\documentclass[a4paper,11pt]{article}
\pdfoutput=1 

\usepackage{jcappub} 

\usepackage[T1]{fontenc} 

\title{\boldmath Adiabatic expansion of polytropic universe with varying cosmological constant: Models tested with observational data}

\author[a]{Ahmet M. Oztas,}
\author[a,b,1]{Emre Dil,\note{Corresponding author.}}
\author[c]{Elif Dil}
\author[d]{and Michael L. Smith}

\affiliation[a]{Department of Physics Engineering, Hacettepe University, TR-06800 Ankara, Turkey}
\affiliation[b]{Department of Physics, Sinop University, TR-57000 Korucuk, Sinop, Turkey}
\affiliation[c]{Department of Statistics, Hacettepe University, TR-06800 Ankara, Turkey}
\affiliation[d]{EVAC, Gilbert, AZ 85214, USA}

\emailAdd{oztas@hacettepe.edu.tr}
\emailAdd{emredil@sakarya.edu.tr}
\emailAdd{elifdil@hacettepe.edu.tr}
\emailAdd{mlsmith55@gmail.com}

\abstract{We use two large collections of observational data of supernovae type Ia (SNe Ia) to investigate the polytropic Universe including the situation with a varying cosmological constant; details of our new derivations are presented. We examine the fitness of our new models of a polytropic Universe with two sets of SNe Ia data to test if these are better descriptions than the current standard model of cosmology. Beginning with the established relationships for polytropic matter we derive new equations describing the influence of polytropic matter on the expanding Universe including the situation with a varying cosmological constant. When the models derived here are tested with large sets of supernovae type I a data we find a significant  influence of polytropic matter on the state of our expanding Universe. We find that one of our models with a varying $\Lambda$ describes the SNe Ia data significantly better than the, $\Lambda$CDM, standard model.}

\begin{document}
\maketitle
\flushbottom

\section{Introduction}
\label{sec:level1}

For the past twenty years, the physicst notice the importance of the cosmological constant $\Lambda$ due to the results of new astronomical investigations \citep{Kirshner,Sola,Bass,Arun}. Later on, the cosmological constant is assumed to be one of the prospects of the dark energy term being the main cause of the accelerated expansion of the universe announced by the observation groups using orbiting satellites such as the Cosmic Background Explorer (COBE) from the cosmic microwave background (CMB) \citep{Dwek}, the WMAP \citep{Komatsu} and the PLANCK mission from the microwave and infrared spectral regions \citep{Planck1}. Finally, the astronomers reached the observational evidence for the cosmological constant with the analysis of supernovae type Ia (SNe Ia) data. The analysis showed that the intervening of the cosmological constant is better fitted to the Friedmann-Robertson-Walker (FRW) model \citep{Riess1,Perlmutter}. However, the data analysis of the PLANCK group has the shortcomings via the requirement of many nuisance parameters, simultaneously, which implies the estimates of the Hubble constant and density parameters is effected from the fitting errors.

As a result of the mentioned disabilities of the standard $\Lambda$CDM model on fitting the SNe Ia, CMB and SDSS data, the physicsits attempted to variate the physical interpretation and mathematical structure of the $\Lambda$CDM model to fix these problems \citep{Melia1,Hartnett,Oztas3,Oztas2,Racz}. One of these problems is to estimate the $H_0$ from SNe Ia, Planck satellite and the SDSS  data \citep{Bennett, Planck3}. While the SNe Ia data determines the value of The Hubble constant about 73-74 km s$^{-1}$ Mpc$^{-1}$, PLANCK group announces the value about 68-70 \citep{Planck3}. In addition, this value is estimated to be about 72-74 km s$^{-1}$ Mpc$^{-1}$ by another investigation with SNe Ia and Cepheid variable star data \citep{Aubourg}. The reson of different estimations on $H_0$ is considering different amount and interactions of constituents and complications of Universe \citep{Riess3}.

In the community, it has been expressed more concretely that the cosmological models for explaining the observed universe should be stricly modified for the past two decades \citep{Carroll1,Carroll2}. One of these modifications methods have been proposed by us \citep{Oztas,Oztas3} and others \citep{Vish,Chen} with an evolving cosmological constant and a modified equation of state (EoS) parameter, $\omega$ \citep{Linder,Carroll3,Barboza,Shen,Sola2}. These modified and complicated models seem to be successfull on estimating better solutions for $H_0$, and  density parameters, and explaining the cosmological constant and dark energy, and in addition to arrive the best fit to observational data.

Inspired by the solution of Lane-Emden equation for a self-gravitating polytropic fluid \citep{Lane}, it is natural to think of the material component of the universe as a fluid in a polytropic adiabatic precess. Here we use the tools of polytropic solutions to estimate the influences of polytropic matter and radiation densities on the nature of our Universe expansion. These more general solutions allow us to simultaneously better estimate the distance of far distant galaxies with the values of normalized matter densities, the cosmological constant and possible spacetime curvature. To this aim we first derive two different solutions of the Friedmann-Robertson-Walker (FRW) model with a varying cosmological constant, $\Lambda$, and contrast these with the current standard model of cosmology, $\Lambda$CDM, with a fixed cosmological constant.

Beginning with the established equation of state for a polytropic system we derive FRW variants, one for a $\Lambda$ being energy density dependent for our expanding Universe and another solution with $\Lambda$ being dependent on the square of Hubble constant. We think both situations possible as these correlate with the history of our expanding Universe. We then cast both new solutions in terms of the normalized parameters; $\Omega_p$, $\Omega_{\Lambda}$ and $\Omega_k$. These solutions allow our derivations to be evaluated with large collections of supernovae data.

Because we have not introduced and extra dimension or field we can check the fitness of our models by analyzing the data from two supernovae type Ia (SNe Ia) collections. The older collection, the Gold data, was published by the group headed by \cite{Riess2} and the second, more recently released, is the larger Union2.1 collection \citep{Suzuki}. We use the tabulated distance modulus, with associated error as a function of the redshift with robust regression for model comparison.

\section{Aidabatic expansion of polytropic Universe}
\label{sec:level2}

Regarding the perfect fluid constituent of the universe as a polytropic fluid, we need to solve the Lane-Emden equation being a version of the Poisson's equation for the spherically-symmetric polytropic fluid. The solution of equation is obtained by the polytropic equation of state of the fluid. The solution describes the evolution of pressure and energy density of the fluid with the polytropic index $\gamma$. Considering the polytopic fluid in isothermal process leads to Chandrasekhar equation \citep{Chandra2}.

We consider an expanding universe insisting of an adiabatic polytropic fluid and a varying cosmological constant. An adiabatic polytropic process is defined by the equation $pV^{\gamma}=Constant$ where $p$ is the fluid pressure and $\gamma=C_p/C_V$. We assume the constituents of the universe are polytropic fluid materials, radiation and the cosmological constant $\Lambda$. The equation of state for the adiabatic polytropic fluid is presented in the Appendix \ref{AppA} as

\begin{equation}
p=K\rho^{\gamma} \,.
\label{a}
\end{equation}
The explicit values of $K$ and $\gamma$ can be obtained as in the Appendix \ref{AppB}. We then proceed by writing the two Friedmann equations

\begin{equation}
\left(\frac{\dot{a}}{a}\right)^2=\frac{8\pi G}{3}\rho+\frac{\Lambda}{3}-\frac{k}{R_0^2a^2} \,,
\label{1}
\end{equation}
\begin{equation}
\frac{\ddot{a}}{a}=-\frac{4\pi G}{3}\left(\rho+3p\right)+\frac{\Lambda}{3} \,.
\label{2}
\end{equation}
Here the cosmological constant is assumed to be time dependent $\Lambda=\Lambda(t)$, G is Newton's gravitational constant, $a$ is the expansion factor, $R$ is our universe radial distance and $\rho$ and $p$ are the energy density and pressure of the polytropic fluid content plus radiation, respectively. Note that $\dot{a}$ and $\ddot{a}$ are the first and second derivatives of the expansion factor with respect to time. 

Taking the derivative of \eqref{1} with respect to time gives

\begin{equation}
\frac{\ddot{a}}{a}=\frac{4\pi G}{3}\left(2\rho+\dot{\rho}\frac{a}{\dot{a}}\right)+\frac{\dot{\Lambda}}{6}\frac{a}{\dot{a}}+\frac{\Lambda}{3} \,.
\label{3}
\end{equation}
Equating both right hand sides of \eqref{2} and \eqref{3} yields to

\begin{equation}
\dot{\rho}=-3\frac{\dot{a}}{a}\left(\rho+K\rho^{\gamma}\right)-\frac{\dot{\Lambda}}{8\pi G} \,,
\label{4}
\end{equation}
where we have replaced \eqref{a} with $p$. It is assumed that the polytropic material contributes to the pressure of the universe with the radiation component and the relation between the pressure and correspondingly  the energy density of polytropic matter and radiation is obtained by following the steps in Appendix \ref{AppB} using the definition of polytropic equation of states for adiabatic (isentropic) process in Appendix \ref{AppA}. Since equation (B.6) is the total pressure of polytropic matter and radiation, the energy density and pressure in the Friedmann equations describe the total values for matter and radiation constituents.

In order to find the exact solution of energy conservation equation \eqref{4}, we need to consider the explicit time dependency of cosmological constant $\Lambda$ for two possible cases.

\subsection{Energy density dependency}
\label{sec:level3}

We use $\Lambda=\textbf{A}\rho$ in \eqref{4} to obtain

\begin{equation}
\dot{\rho}=-3\frac{\dot{a}}{a}\left(\rho+K\rho^{\gamma}\right)-\textbf{A}\frac{\dot{\rho}}{8\pi G} \,.
\label{5}
\end{equation}
Rearranging \eqref{5} and integrating both sides presents us with

\begin{equation}
\left(1+\frac{\textbf{A}}{8\pi G}\right)\int{\frac{d\rho}{\rho+K\rho^{\gamma}}}=-3\int{\frac{da}{a}} \,.
\label{6}
\end{equation}
We express the integrand in a rational form as

\begin{equation}
\left(1+\frac{\textbf{A}}{8\pi G}\right)\frac{1}{\gamma-1}\int{\left[\frac{\gamma}{\rho}-\frac{1+\gamma K\rho^{\gamma-1}}{\rho+K\rho^{\gamma}}\right]d\rho}=\ln{\frac{C}{a^3}} \,,
\label{7}
\end{equation}
where $C$ is the integration constant. Integrating the left side of  \eqref{7} and rearranging leads to

\begin{equation}
\ln{\left(\frac{\rho^{\gamma}}{\rho+K\rho^{\gamma}}\right)}=\ln{\left(\frac{C}{a^3}\right)}^{\left(\gamma-1\right)M} \,,
\label{8}
\end{equation}
where our M represents not mass but the term

\begin{equation}
M=\frac{8\pi G}{A+8\pi G} \,.
\label{9}
\end{equation}
Solving \eqref{8} for the matter density yields

\begin{equation}
\rho=\frac{C^M}{\left[a^{3\left(\gamma-1\right)M}-KC^{\left(\gamma-1\right)M}\right]^{\frac{1}{\gamma-1}}} \,.
\label{10}
\end{equation}

By using the present day values of $t=t_0$ and $\rho=\rho_0$ and $a(t_0)=1$ in \eqref{10}, we can solve for $C$ as
\begin{equation}
C^{\left(\gamma-1\right)M}=\frac{\rho_0^{\gamma-1}}{1+K\rho_0^{\gamma-1}} \,,
\label{11}
\end{equation}
and inserting this into \eqref{10} we find the energy density

\begin{equation}
\rho=\frac{\rho_0}{\left[a^{3\left(\gamma-1\right)M}\left(1+K\rho_0^{\gamma-1}\right)-K\rho_0^{\gamma-1}\right]^{\frac{1}{\gamma-1}}} \,.
\label{12}
\end{equation}
We now rearrange \eqref{1} in which energy density \eqref{12} is used to obtain the luminosity distance $D_L$ as

\begin{equation}
\left(\frac{da}{dt}\right)^2=a^2\left(\frac{8\pi G}{3}\rho+\frac{\Lambda}{3}-\frac{k}{R_0^2a^2}\right) \,,
\label{13}
\end{equation}
and by taking square root and rearranging \eqref{13} gives us

\begin{equation}
dt=\frac{da}{a\sqrt{\frac{8\pi G}{3}\rho+\frac{\Lambda}{3}-\frac{k}{R_0^2a^2}}} \,.
\label{14}
\end{equation}
We use the normalized density parameters beginning with  $\Omega_i=\rho_i/\rho_c$, where $\rho_c$ is the critical energy density and $\rho_i$ is the energy density of different matter species, for instance baryons and neutrinos. We continue with the list of commonly used normalized parameters as

\begin{equation}
\Omega_p=\frac{\rho_0}{\rho_c} \,,
\qquad
\Omega_{\Lambda}=\frac{\Lambda}{3H_0^2} \,,
\qquad
\Omega_k=-\frac{k}{R_0^2H_0^2} \,,
\label{15}
\end{equation}
for the polytropic fluid, cosmological constant and space-time curvature, respectively. $\Omega_p$ stands for the total density parameter of matter and radiation components in terms of the polytropic matter density. The contribution of each part is described by the ratio parameter $\beta$. Here, the critical density is $\rho_c=3H_0^2/8\pi G$ and $\gamma=4/3$ for the polytropic fluid. We then substitute \eqref{12} and \eqref{15} into \eqref{14} to yield

\begin{equation}
dt=\frac{da}{H_0a\left[\frac{8\pi G}{3H_0^2}\frac{\rho_0}{\left(a^M(1+K\rho_0^{1/3})-K\rho_0^{1/3}\right)^3}+\frac{\Lambda}{3H_0^2}-\frac{k}{a^2R_0^2H_0^2}\right]^{1/2}} \,,
\label{16}
\end{equation}
and after substitution with the normalized parameters we get

\begin{equation}
dt=\frac{a^{(3M-2)/2}da}{H_0\sqrt{\frac{\Omega_p}{\textbf{A}_p^3}+\Omega_{\Lambda}a^{3M}+\Omega_ka^{3M-2}}} \,,
\label{17}
\end{equation}
where $\textbf{A}_p=1+K\rho_0^{1/3}-(K\rho_0^{1/3}/a^M)$. 

From the radial null geodesic condition of the FRW metric, we obtain

\begin{equation}
\frac{dt}{a}=R_0\frac{dr}{\sqrt{1-kr^2}} \,,
\label{18}
\end{equation}
and after inserting  this into \eqref{17} we must integrate both sides to get

\begin{equation}
\int_0^r{\frac{H_0R_0dr}{\sqrt{1+r^2R_0^2H_0^2\Omega_k}}}=\int_a^1{\frac{a^{(3M-4)/2}da}{\sqrt{\frac{\Omega_p}{A_p^3}+\Omega_{\Lambda}a^{3M}+\Omega_ka^{3M-2}}}} \,,
\label{19}
\end{equation}
where we have used $k=-R_0^2H_0^2\Omega_k$. 

We must now change the variable $y=\sqrt{\Omega_k}R_0H_0r$ on the left hand side of integral \eqref{19} leading to

\begin{equation}
D_L=\frac{c}{H_0a\sqrt{|\Omega_k|}}\texttt{sinn}{\left(\sqrt{|\Omega_k|}\int_a^1{\frac{a^{(3M-4)/2}da}{\sqrt{\frac{\Omega_p}{A_p^3}+\Omega_{\Lambda}a^{3M}+\Omega_ka^{3M-2}}}}\right)} \,.
\label{20}
\end{equation}
Here, $D_L=R_0r/a$ is used and $\texttt{sinn}$ refers to $\sin$ for negative spacetime curvature and to $\sinh$ for positive spacetime curvature. We also need to change the variable in the integral of \eqref{20} from the scale factor $a$ to redshift $z$ using $a=1/(1+z)$ so that we arrive at

\begin{equation}
D_L=\frac{c(1+z)}{H_0\sqrt{|\Omega_k|}}\texttt{sinn}{\left(\sqrt{|\Omega_k|}\int_0^z{\frac{dz}{\sqrt{\frac{\Omega_p (1+z)^{3M}}{A_p^3}+\Omega_{\Lambda}+\Omega_k(1+z)^2}}}\right)} \,.
\label{21}
\end{equation}
For a flat universe with $\Omega_k=0$ equation \eqref{21} takes the form

\begin{equation}
D_L=\frac{c(1+z)}{H_0}\int_0^z{\frac{dz}{\sqrt{\frac{\Omega_p}{A_p^3}(1+z)^{3M}+\Omega_{\Lambda}}}} \,.
\label{22}
\end{equation}
The term $M$ in \eqref{9} can also be obtained from the value of $\textbf{A}$, where $\Lambda=\textbf{A}\rho$ and

\begin{equation}
\textbf{A}=8\pi G\frac{\Omega_{\Lambda}}{\Omega_p} \Longrightarrow M=\frac{\Omega_p}{\Omega_p +\Omega_{\Lambda}} \,.
\label{23}
\end{equation}
For a flat universe $M=\Omega_p$ since $\Omega_p+\Omega_{\Lambda}=1$, and for a curved universe $M=\Omega_p/(1-\Omega_k)$ since $\Omega_p+\Omega_{\Lambda}+\Omega_k=1$.

\subsection{Hubble parameter dependency}
\label{sec:level4}

For the Hubble parameter dependency of the cosmological constant we use $\Lambda=3\textbf{A}H^2=3\textbf{A}(\dot{a}/a)^2$ in \eqref{4} and obtain the energy conservation equation as

\begin{equation}
\dot{\rho}=-3\frac{\dot{a}}{a}\left(\rho+K\rho^{\gamma}\right)-\frac{3\textbf{A}}{8\pi G}2\frac{\dot{a}}{a}\left[\frac{\ddot{a}}{a}-\left(\frac{\dot{a}}{a}\right)^2\right] \,.
\label{24}
\end{equation}
The last term in the parentheses is obtained from \eqref{1} and \eqref{2} for a flat universe and is required to derive an analytical solution for energy density of the polytropic fluid as follows

\begin{equation}
\dot{\rho}=-3\left(1-\textbf{A}\right)\frac{\dot{a}}{a}\left(\rho+K\rho^{\gamma}\right) \,.
\label{25}
\end{equation}
This differential equation has the same form as equation \eqref{6} and can be solved by following the same steps as for \eqref{7} and \eqref{8};

\begin{equation}
\rho=\frac{\textbf{C}}{\left[a^{3\left(\gamma-1\right)N}-K\textbf{C}^{\left(\gamma-1\right)}\right]^{\frac{1}{\gamma-1}}} \,,
\label{26}
\end{equation}
where $\textbf{C}$ is the integration constant and $N=1-\textbf{A}$. Using present day $t_0$ values for $a(t_0)=1$ and $\rho=\rho_0$ in \eqref{26}, we obtain the constant $\textbf{C}$ as

\begin{equation}
\textbf{C}^{\left(\gamma-1\right)}=\frac{\rho_0^{\gamma-1}}{1+K\rho_0^{\gamma-1}} \,,
\label{27}
\end{equation}
which is used in equation \eqref{26} to obtain the energy density

\begin{equation}
\rho=\frac{\rho_0}{\left[a^{3\left(\gamma-1\right)N}\left(1+K\rho_0^{\gamma-1}\right)-K\rho_0^{\gamma-1}\right]^{\frac{1}{\gamma-1}}} \,.
\label{28}
\end{equation}
To obtain the luminosity distance $D_L$, we substitute the energy density of equation \eqref{28} into \eqref{14} with $\gamma=4/3$ and $k=0$ for a flat universe, such that

\begin{equation}
dt=\frac{da}{H_0a\left[\frac{8\pi G}{3H_0^2}\frac{\rho_0}{\left(a^N(1+K\rho_0^{1/3})-K\rho_0^{1/3}\right)^3}+\frac{\Lambda}{3H_0^2}\right]^{1/2}} \,,
\label{29}
\end{equation}
and we use the density parameters from equation \eqref{15} in this as

\begin{equation}
dt=\frac{a^{(3N-2)/2}da}{H_0\sqrt{\frac{\Omega_p}{\tilde{A}_p^3}+\Omega_{\Lambda}a^{3N}}} \,,
\label{30}
\end{equation}
where $\tilde{A}_p=1+K\rho_0^{1/3}-(K\rho_0^{1/3}/a^N)$. 

The radial null geodesic condition for the flat FRW metric is

\begin{equation}
\frac{dt}{a}=R_0dr \,.
\label{31}
\end{equation}
Equating both right hand sides of \eqref{30} and \eqref{31} yields

\begin{equation}
\int_0^r{H_0R_0dr}=\int_a^1{\frac{a^{(3N-4)/2}da}{\sqrt{\frac{\Omega_p}{\tilde{A}_p^3}+\Omega_{\Lambda}a^{3N}}}} \,,
\label{32}
\end{equation}
and using $D_L=\frac{R_0r}{a}$ gives us

\begin{equation}
D_L=\frac{c}{H_0a}\int_a^1{\frac{a^{(3N-4)/2}da}{\sqrt{\frac{\Omega_p}{\tilde{A}_p^3}+\Omega_{\Lambda}a^{3N}}}} \,.
\label{33}
\end{equation}
We then convert the integration variable from the scale factor to the redshift as in equation \eqref{21}

\begin{equation}
D_L=\frac{c(1+z)}{H_0}\int_0^z{\frac{dz}{\sqrt{\frac{\Omega_p}{\tilde{A}_p^3}(1+z)^{3N}+\Omega_{\Lambda}}}} \,.
\label{34}
\end{equation}
To find $N=1-\textbf{A}$ we use $\Lambda=3\textbf{A}H^2$ for $t=t_0$, and we get

\begin{equation}
A=\frac{\Lambda}{3H_0^2}=\Omega_{\Lambda} \Longrightarrow N=1-\Omega_{\Lambda}=\Omega_p \,
\label{35}
\end{equation}
and since we are considering a flat universe, $\Omega_p+\Omega_{\Lambda}=1$.

Comparing the luminosity distance, $D_L$, results in \eqref{34} and \eqref{22} for the special case of a flat universe with $M=N=\Omega_p$, we notice that both are identical. So we conclude that the polytropic solutions with varying cosmological constant ($\propto\hskip-1mm\rho$ and $\propto \hskip-1mm H^2$) have the same luminosity distance results - for a flat universe.

\subsection{Non-varying cosmological constant}
\label{sec:level5}

This case was previously studied by \cite{Oztas}, where the derivative of $\Lambda$ with respect to time in equation \eqref{4} vanishes. Therefore, we proceed with $A=\textbf{A}=0$ for equations \eqref{5} and \eqref{24} and the energy density now has the form

\begin{equation}
\rho=\frac{\rho_0}{\left[a^{3\left(\gamma-1\right)}\left(1+K\rho_0^{\gamma-1}\right)-K\rho_0^{\gamma-1}\right]^{\frac{1}{\gamma-1}}} \,.
\label{36}
\end{equation}
Here $M=N=1$ for the $\rho$ and $H^2$ independent $\Lambda$. Therefore, the luminosity distance is obtained as in equation \eqref{21}

\begin{equation}
D_L=\frac{c(1+z)}{H_0\sqrt{|\Omega_k|}}\texttt{sinn}{\left(\sqrt{|\Omega_k|}\int_0^z{\frac{dz}{\sqrt{\frac{\Omega_p(1+z)^{3}}{A_p^3}+\Omega_{\Lambda}+\Omega_k(1+z)^2}}}\right)} \,,
\label{37}
\end{equation}
for a universe with curved spacetime, and as \eqref{22} or \eqref{34}

\begin{equation}
D_L=\frac{c(1+z)}{H_0}\int_0^z{\frac{dz}{\sqrt{\frac{\Omega_p}{\tilde{A}_p^3}(1+z)^{3}+\Omega_{\Lambda}}}} \,.
\label{38}
\end{equation}
for a flat universe.
     
\section{Results: Testing models with astronomical data}
\label{sec:level6}

In the previous section we present the polytropic universe model with a cosmological constant both varying and constant over time. In Section \ref{sec:level3} we propose the cosmological constant varies with energy density $\rho$, while it varies with the square of Hubble parameter $H^2$ in Section \ref{sec:level4} and a non-varying cosmological constant in Section \ref{sec:level5}. We test the usefulness of these model fit using SNe Ia data of the Gold \citep{Riess2,Tonry} and Union2.1 \citep{Suzuki} collections. We use the distance magnitudes and associated errors along with the related redshifts from both collections.

For the flat polytropic universe with a varying cosmological constant, we fit the luminosity distance $D_L$ \eqref{22} and \eqref{34} to both Gold and Union2.1 data. We also fit the $D_L$ value in \eqref{38} to both Gold and Union2.1 data for the flat polytropic universe with a non-varying cosmological constant. Moreover, we use the distance \eqref{21} for a non-flat polytropic universe with a varying cosmological constant, and use \eqref{37} for a non-flat polytropic universe with a non-varying cosmological constant, then fit these distance values with the Gold data only.

To compare the models and obtain the nuissance parameters with the observational data, we perform the regression analysis of $\chi^2$ minimization routine for the $\mu_{th}$ modulus values of our theoretical models and the observational $\mu_{ob}$ values,

\begin{equation}
\chi^2=\sum_{i=1}^{\texttt{data}}{\frac{\left[\mu_{th}(z_i;\Omega_p,\Omega_k,h,\beta)-\mu_{ob,i}\right]^2}{\sigma_i^2}} \,,
\label{39}
\end{equation}
where $\Omega_k$ vanishes for the flat model fits, $h$ is the dimensionless constant in Hubble parameter (H=100h), $\sigma_i$'s are the error values in $\mu_{ob}$ values in the observational data. We totally have four polytropic models from which we obtain the theoretical $\mu_{th}$ values. From equations \eqref{22} and \eqref{34} we calculated modulus values for flat polytropic universe with a varying cosmological constant, and we compare it by both Gold and Union2.1 data in $\chi^2$ minimization procedure. We also determine the $\mu$ value in \eqref{38} for flat polytropic universe with a non-varying cosmological constant, and compare that again by both Gold and Union2.1 data in $\chi^2$ minimization. Furthermore, we calculate $\mu$ modulus for \eqref{21} for a non-flat polytropic universe with a varying cosmological constant, and for \eqref{37} for a non-flat polytropic universe with a non-varying cosmological constant, and compare both modulus by Gold data only. In all $\mu_{th}$ values, we use $D_L$ term in megaparsec (Mpc), such that
\begin{table}
	\centering
	\caption{Polytropic model designations and free parameter numbers}
    	\label{tab0}
	\begin{tabular}{c c c c } 
		\hline
		No. & Designation & Equation & Free parameters \\
		\hline
		1 & Flat-varying $\Lambda$ & \eqref{34} & 3 \\
		2 & Flat-Const $\Lambda$ & \eqref{38} & 3 \\
		3 & Curved-Varying $\Lambda$ & \eqref{21} & 4 \\
        4 & Curved-Const $\Lambda$  & \eqref{37} & 4 \\
		\hline   
	\end{tabular}
\end{table}

\begin{table}
	\centering
	\caption{Random fit of parameters from $\chi^2$ minimization for Gold Data.}
    	\label{tab1}
	\begin{tabular}{c c c c c c c c} 
		\hline
		No. & $\Omega_P$ & $\Omega_k$ & h & $\beta$ & $\chi^2$ & BIC & AIC\\
		\hline
		1 & 0.67 & - & 0.694 & 0.73 & 351 & 366 & 357\\
		2 & 0.28 & - & 0.692 & 0.86 & 325 & 340 & 330\\
		3 & 1.27 & -0.86 & 0.689 & 0.71 & 318 & 338 & 326\\
        4 & 1.03 & -0.36 & 0.655 & 0.83 & 380 & 400 & 388\\
		\hline
	\end{tabular}
\end{table}

\begin{table}
	\centering
	\caption{Prior fit of parameters from $\chi^2$ minimization for Gold Data.}
	\label{tab2}
	\begin{tabular}{c c c c c c c c} 
		\hline
		No. & $\Omega_P$ & $\Omega_k$ & h & $\beta$ & $\chi^2$ & BIC & AIC\\
		\hline
		1 & 0.29 & - & 0.758 & 0.99 & 464 & 469 & 465 \\
		2 & 0.29 & - & 0.703 & 0.99 & 340 & 345 & 342 \\
		3 & 0.27 & 0.02 & 0.751 & 0.99 & 469 & 474 & 471 \\
        4 & 0.27 & 0.02 & 0.728 & 0.99 & 500 & 505 & 502 \\
		\hline
	\end{tabular}
\end{table}

\begin{table}
	\centering
	\caption{Random fit of parameters from $\chi^2$ minimization for Union2.1 Data.}
	\label{tab3}
	\begin{tabular}{c c c c c c c c} 
		\hline
		No. & $\Omega_P$ & h & $\beta$ & $\chi^2$ & BIC & AIC\\
		\hline
		1 & 0.53 & 0.747 & 0.83 & 2069 & 2087 & 2074 \\
		2 & 0.29 & 0.753 & 0.96 & 2197 & 2216 & 2203 \\
		\hline
	\end{tabular}
\end{table}

\begin{table}
	\centering
	\caption{Prior fit of parameters from $\chi^2$ minimization for Union2.1 Data.}
	\label{tab4}
	\begin{tabular}{c c c c c c c c} 
		\hline
		No. & $\Omega_P$ & h & $\beta$ & $\chi^2$ & BIC & AIC\\
		\hline
		1 & 0.29 & 0.775 & 0.99 & 2183 & 2190 & 2185 \\
		2 & 0.29 & 0.754 & 0.99 & 2197 & 2204 & 2199 \\
		\hline
	\end{tabular}
\end{table}

\begin{equation}
\mu_{th}=5\log{\left(\frac{D_L}{Mpc}\right)}+25 \,,
\label{40}
\end{equation}
where $D_L$ is a function of nuissance parameters of $(\Omega_p,\Omega_k,h,\beta)$ for non-flat models in \eqref{21} and \eqref{37}, and of $(\Omega_p,h,\beta)$ for flat models in \eqref{22}, \eqref{34} and \eqref{38}.

After designating the model abbreviation in Table \ref{tab0}, we list regression parameters as the fit values for $(\Omega_p,h,\beta)$ and $\chi^2$ values for flat models, and for $(\Omega_p,\Omega_k,h,\beta)$ and $\chi^2$ values for non-flat models in Table \ref{tab1} and Table \ref{tab3} for the Gold Data and Union2.1 Data, respectively. Since the model function is too complex to be optimized by classical optimization algorithms, the minimization process fall into the local optimum many times. Therefore, we applied meta-heuristic optimization methods, such as 'genetic algorithm' and 'simulated annealing algorithm', to minimize the $\chi^2$ values for our four models. These random fitting minimization routines fall into the local optimum, and we therefore obtained the fitted functions by using some non-informative priors as the fixed regression parameters in classical optimization algorithm. The prior used fitting values for the variables $(\Omega_p,\Omega_k,h,\beta)$ are listed in Table \ref{tab2} and Table \ref{tab4} for Gold and Union2.1 Data, respectively, with the corresponding $\chi^2$ values.

In order to find the best model, we perform the statistical performance evaluations; Bayesian Information Criteria (BIC) and Akaike Information Criteria (AIC). The BIC is defined as $-2\ln{L}+k\ln{n}$, AIC is defined as $-2\ln{L}+2k$ where $n$ is the number of data pairs, $k$ number of parameters and $L$ is the likelihood function, such that

\begin{equation}
L=\prod_i{\frac{1}{\sqrt{2\pi}\sigma_i}\exp{\left[-\frac{\left(\mu_{th}(z_i;\Omega_p,\Omega_k,h,\beta)-\mu_{ob,i}\right)^2}{2\sigma_i^2}\right]}} \,.
\label{41}
\end{equation}
Since the exponent term in $L$ is the summand in the $\chi^2$, BIC and AIC values can be obtained easily from $\chi^2$ values for each of the six results, which are listed in Table \ref{tab1}-\ref{tab4}. According to the information criteria values, 3rd model NFlat-Varying has the best random fit values with least AIC-BIC in Table \ref{tab1}, 2nd model Flat-Const is best function with least AIC-BIC values for the prior used minimization in Table \ref{tab2} for Gold Data. Moreover, 1st Flat-Varying model has the best random and prior fit values in Tables \ref{tab3} and Table \ref{tab4} for Union2.1 Data. After the best model test with $\chi^2$, we study the curve fit of each model independently in order to investigate whether each single model is accurate, or not?

\begin{table}
	\centering
	\caption{Statistical confidence values for Gold Data.}
	\label{tab5}
	\begin{tabular}{c c c c c c} 
		\hline
	   No.& $r^2$  &  FitStd.Err. &  Fstat \\
     \hline
   2& 0.9931  & 0.3179 & 22620 \\
   4& 0.9929  & 0.3223 & 22000 \\
   1& 0.9896  & 0.3905 & 14940 \\
   3& 0.9893  & 0.3955 & 14560 \\
		\hline
	\end{tabular}
\end{table}

\begin{table}
	\centering
	\caption{Statistical confidence values for Union2.1 Data.}
	\label{tab6}
	\begin{tabular}{c c c c c c} 
		\hline
	  No.&  $r^2$  &  FitStd.Err. &  Fstat \\
     \hline
   2& 0.9857  & 0.3821 & 40030 \\
   1& 0.9833  & 0.4126 & 34250 \\
		\hline
	\end{tabular}
\end{table}

\begin{figure}
	\includegraphics[width=\columnwidth]{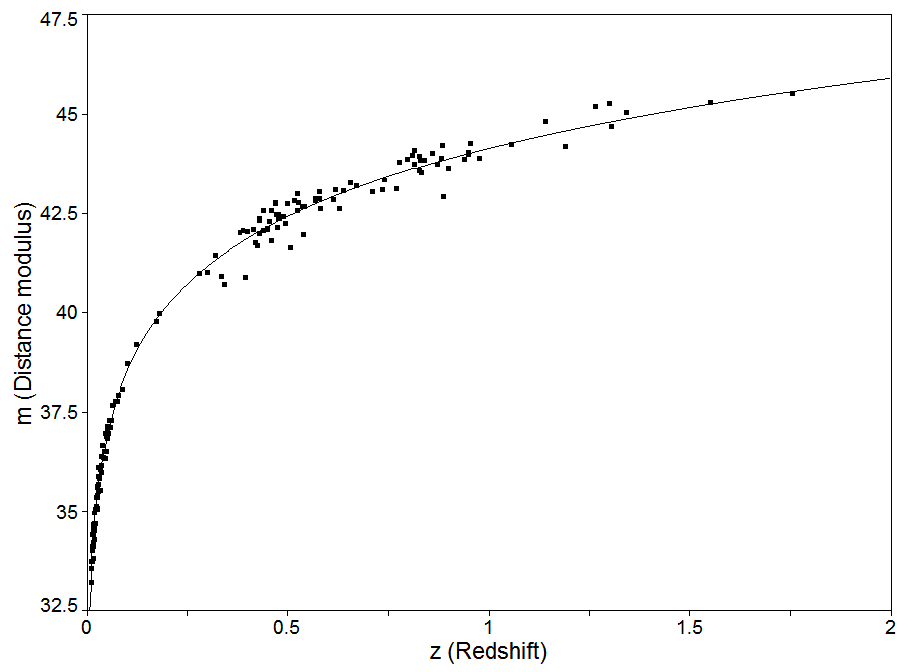}
    \caption{Hubble diagram for Gold Data.}
    \label{fig1}
\end{figure}

\begin{figure}
	\includegraphics[width=\columnwidth]{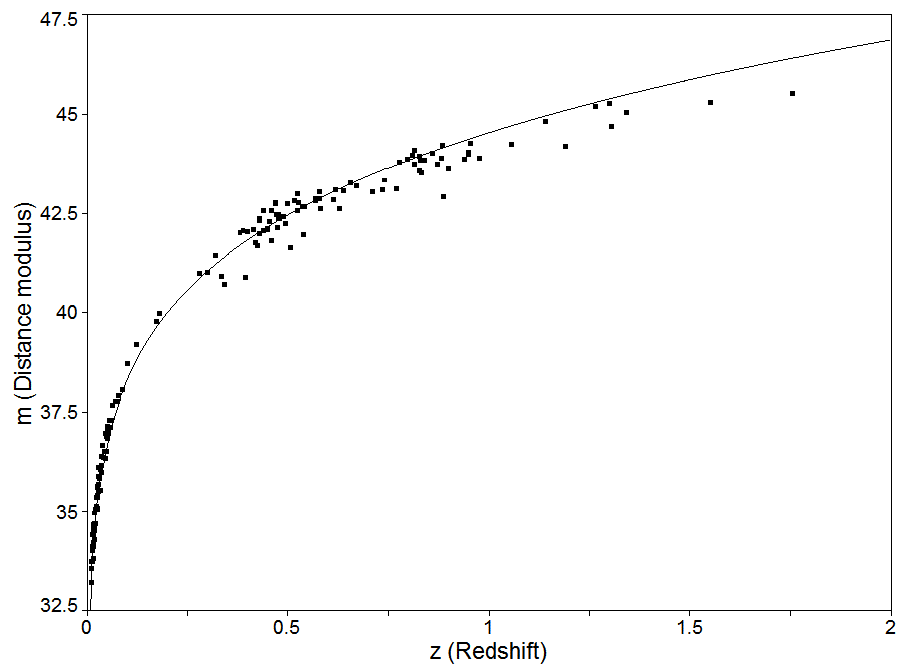}
  \caption{Hubble diagram fit for polytropic flat universe with varying cosmological constant model according to Gold Data.}
    \label{fig2}
\end{figure}

\begin{figure}
	\includegraphics[width=\columnwidth]{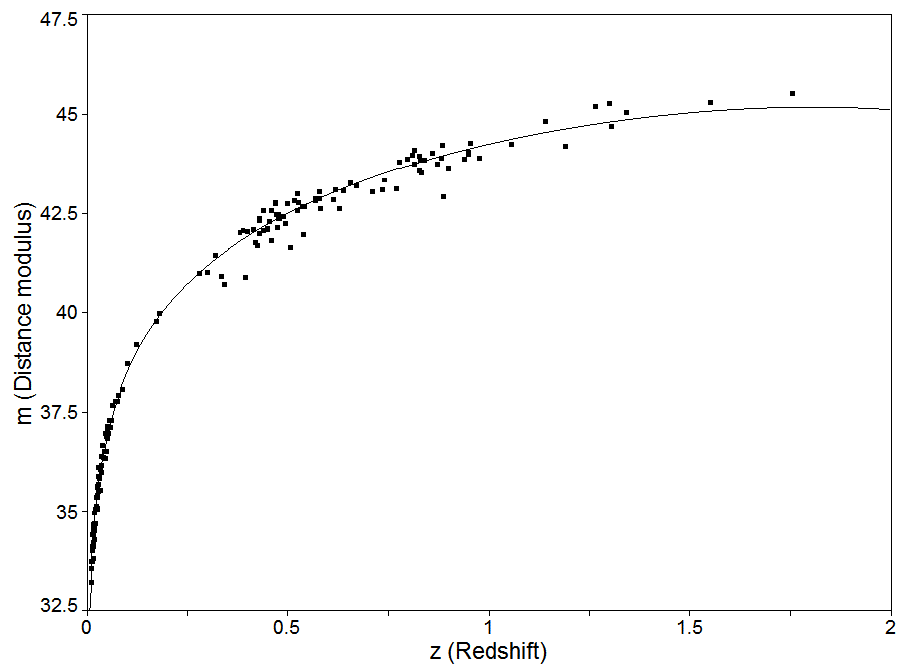}
  \caption{Hubble diagram fit for polytropic flat universe with non-varying cosmological constant model according to Gold Data.}
    \label{fig3}
\end{figure}

\begin{figure}
	\includegraphics[width=\columnwidth]{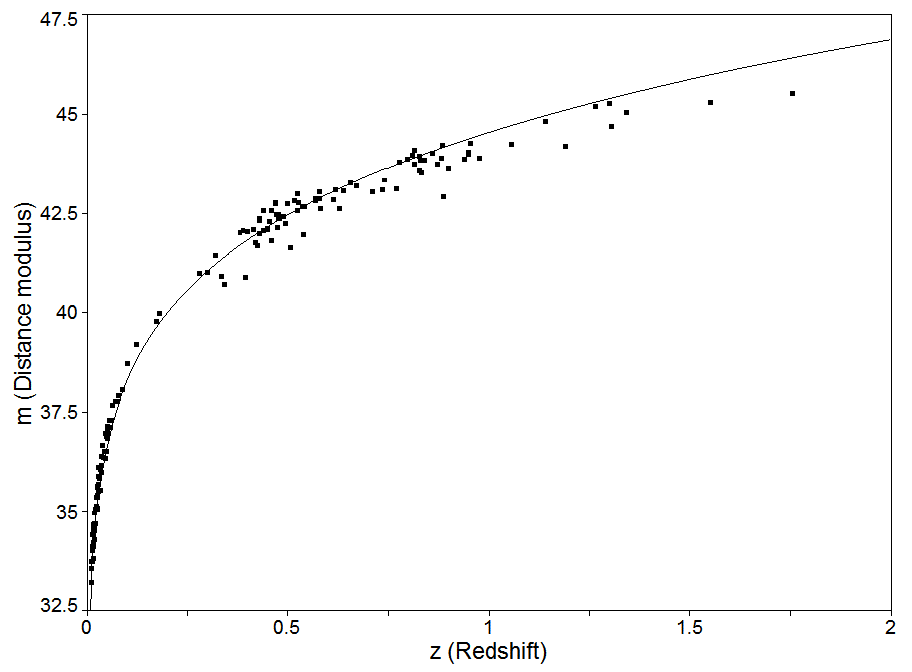}
  \caption{Hubble diagram fit for polytropic non-flat universe with varying cosmological constant model according to Gold Data.}
    \label{fig4}
\end{figure}

\begin{figure}
	\includegraphics[width=\columnwidth]{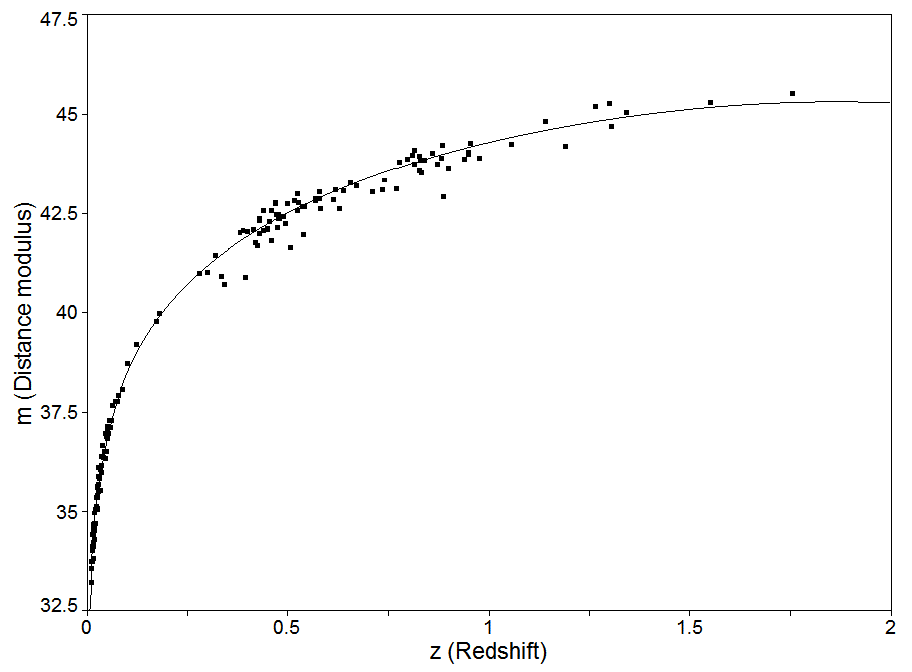}
  \caption{Hubble diagram fit for polytropic non-flat universe with non-varying cosmological constant model according to Gold Data.}
    \label{fig5}
\end{figure}

\begin{figure}
	\includegraphics[width=\columnwidth]{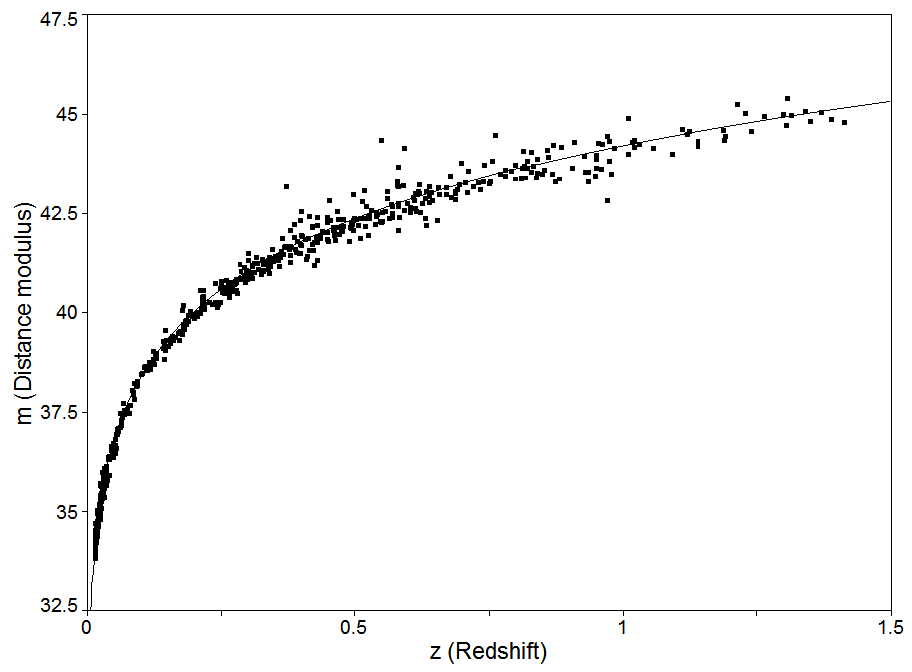}
    \caption{Hubble diagram for Union2.1 Data.}
    \label{fig6}
\end{figure}

\begin{figure}
	\includegraphics[width=\columnwidth]{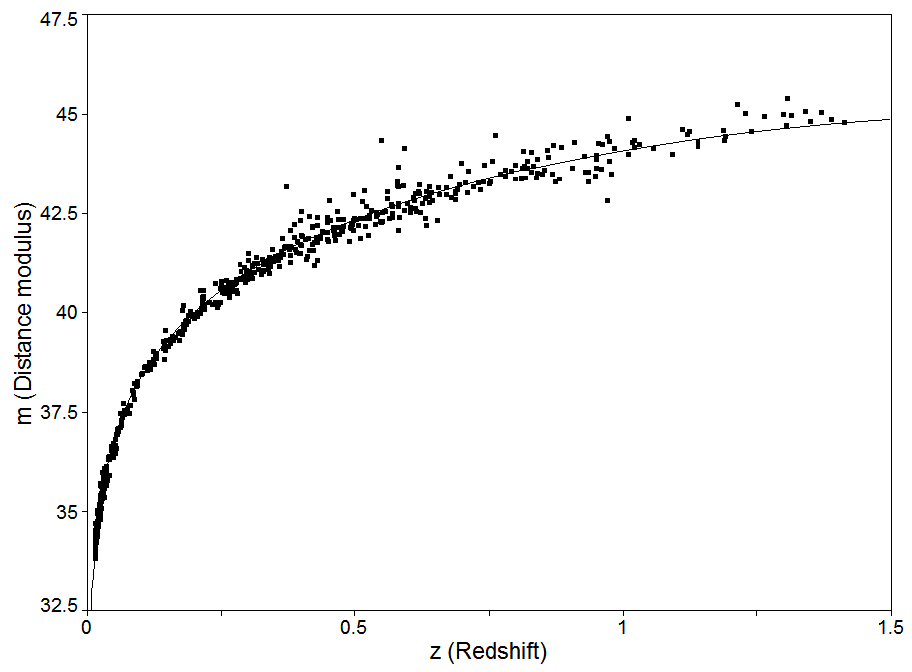}
  \caption{Hubble diagram fit for polytropic flat universe with varying cosmological constant model according to Union2.1 Data.}
    \label{fig7}
\end{figure}

\begin{figure}
	\includegraphics[width=\columnwidth]{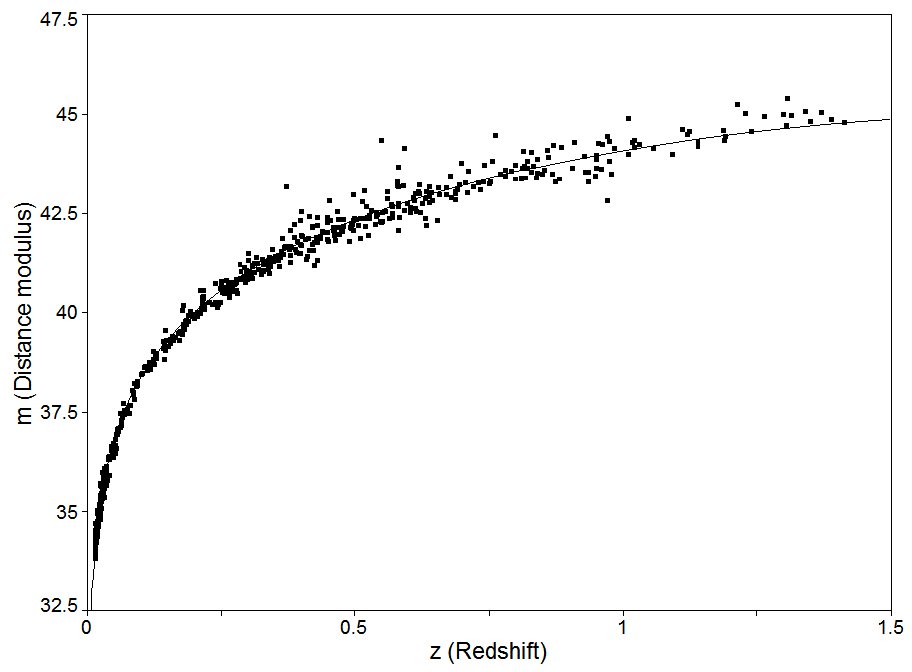}
  \caption{Hubble diagram fit for polytropic flat universe with non-varying cosmological constant model according to Union2.1 Data.}
    \label{fig8}
\end{figure}

We present the diagrams for the observed and estimated distance modulus versus redshift values for our four polytropic models. Observational Hubble diagram for Gold data is illustrated in Fig. \ref{fig1}, while the polytropic flat and non-flat model diagrams are presented in Figs. \ref{fig2}-\ref{fig5} for varying and non-varying cosmological constant. Moreover, observational Union2.1 data Hubble diagram is presented in Fig. \ref{fig6}, with flat varying and non-varying cosmological constant models are sketched in Figs. \ref{fig7} and \ref{fig8}, respectively.

We infer from the statistical determination coefficient $r^2$ and error values in Table \ref{tab5} and Table \ref{tab6} that all the models are accurate with the observational data. These $r^2$ values are for the statistical measures of how close the data are to the fitted model, and 100 percent indicates that the model explains data perfect. Fstat test (Anova) which is the test for the inaccuracy of the model gives $p-value=0.000$ for all models. Therefore, we could interpret that all the models are accurate with 95 percent confidence.

\section{Conclusions}

We consider a FRW universe consisting of a polytropic matter constituent with the associating radiation part being the solution of Lane-Emden equation, and the cosmological constant constituent. We investigate both flat and curved FRW universe with a varying and non-varying cosmological constant cases. Totally, we consider four polytropic models designated in Table \ref{tab0}. The flat FRW universe consisting of the polytropic material and a varying cosmological constant corresponds to Model 1 and a non-varying cosmological constant corresponds to Model 2. On the other hand, the non-flat (curved) FRW universe consisting of the polytropic material and a varying cosmological constant corresponds to Model 3 and a non-varying cosmological constant corresponds to Model 4.

For each model, we first set the Friedmann equations and solve them to obtain the total energy densities. Then, we proceed by determining the luminosity distances for the proposed methods from the corresponding energy densities and Friedmann equations. These distance values are used to obtain the theoretical modulus values $\mu_{th}$ for which we perform the data analysis with the observational Gold and Union2.1 data sets. Since these data sets include the observational modulus values and associated errors in $\mu_{ob}$, we could perform the $\chi^2$ mininization process in order to find the best fit model with the prior and fit parameters in Tables \ref{tab1}-\ref{tab4}. The minimzations procedure yields the best fit model as the Model 2 for the Gold data, and Model 1 for the Union2.1 data sets. 

These results show that the flatness assumption for the geometry of universe can be accepted as a better estimation on the shape of Universe. Also constant $Lambda$ model in a flat polytropic universe are more consistent with the Gold data observations, while the varying $\Lambda$ model are in Union2.1 data. This fact presents that the recent observations of Suzuki's group supports the idea of a varying cosmological constant instead of a non-varying cosmological constant idea. This may cause due to the improving measurement accuracy of the recent observational instruments.

Also, we evaluate the model accuracy to the data sets by performing the $r^2$ and Anova tests while results are given in Tables \ref{tab5} and \ref{tab6} for gold and Union2.1 data sets. According to these results we find all four polytropic models are consistent with the data.

These results imply that considering the constituents of the universe as a polytropic matterial with the associating radiation part and expressing the total pressure and density of the polytropic material and radiation in terms of the matter leads to a convenient scheme for the flat and curved FRW universes with a varying cosmological constant. In addition, our Model 2 with a $\chi^2$ of 340 in \ref{tab2} presents a better fit than the standard $\Lambda$CDM model with a $\chi^2$ of 623 for $\Omega_p=0.29$ and $h=0.75$ for the Gold Data.

We hope these alternatives interpretations on the nature of our universe in terms of the species of constituents and the behavior of the constituents will provide a better understanding of the physics underlying the cosmos, and a better estimation of cosmological parameters for the astronomers and physicists.

\appendix

\section{Adiabatic polytropic equation}
\label{AppA}

We derive the the equation of state for a fluid in an adiabatic polytopic process. The specific heat coefficients are given as

\begin{equation}
\gamma=\frac{C_p}{C_V} \,,
\qquad
C_p-C_V=R \,,
\label{b}
\end{equation}
where $C_p$ and $C_V$ are the specific heat at constant pressure and volume, respectively, and $R$ is the gas constant. Using both equations in \eqref{b} gives

\begin{equation}
\frac{C_p}{R}=\frac{\gamma}{\gamma-1} \,.
\label{c}
\end{equation}
Also the equation of state for an ideal gas is given as

\begin{equation}
p=\rho R T \,.
\label{d}
\end{equation}
where $\rho$ is the density, and $T$ is the temperature. The entropy of the gas is given by:

\begin{equation}
dS=\frac{dQ}{T}=C_p\frac{dT}{T}-R\frac{dp}{p} \,,
\label{e}
\end{equation}
Since $dQ$ is differential change in heat, for an adiabatic process it is zero and therefore

\begin{equation}
dS=0 \Longrightarrow C_p\frac{dT}{T}=R\frac{dp}{p} \,.
\label{f}
\end{equation}
Substituting \eqref{d} in \eqref{f} yields

\begin{equation}
C_pdT=\frac{dp}{\rho} \Longrightarrow \frac{C_p}{R}d\left(\frac{p}{\rho}\right)=\frac{dp}{\rho} \,,
\label{g}
\end{equation}
Differentiating \eqref{g} gives

\begin{equation}
\left(\frac{C_p}{R}-1\right)\frac{dp}{p}=\frac{C_p}{R}\frac{d\rho}{\rho} \,.
\label{h}
\end{equation}
Inserting \eqref{c} into \eqref{h} leads to

\begin{equation}
\frac{1}{\gamma-1}\frac{dp}{p}=\frac{\gamma}{\gamma - 1}\frac{d\rho}{\rho} \,,
\label{i}
\end{equation}
which reduces to

\begin{equation}
\frac{dp}{p}=\gamma\frac{d\rho}{\rho} \,.
\label{j}
\end{equation}
When we integrate \eqref{j}, we finally obtain the equation of state for a fluid in an adiabatic polytorpic process:

\begin{equation}
p=K\rho^{\gamma} \,.
\label{k}
\end{equation}

\section{Pressure of polytropic fluid}
\label{AppB}

It is assumed that the total pressure of the Universe is due to matter, radiation and cosmological constant. The pressure of the constituents other than the cosmological constant is defined as

\begin{equation}
p_m=\frac{k}{\mu \textbf{H}}\rho T \,,
\label{l}
\end{equation}
where $k$ is the Boltzmann contant, $\mu$ is the mean molecular weight which is taken 1 under normal conditions, $\textbf{H}$ is the mass of the hydrogen atom, and $T$ is the absolute temperature. Since it is very small as 2.73 K, the pressure of the matter content is considered to be negligible (Caroll).But in the polytopic universe model, we do not neglect it. The radation pressure is defined to be

\begin{equation}
p_{rad}=\frac{1}{3}\alpha T^4 \,,
\label{m}
\end{equation}
where $\alpha$ is the Stefan's radiation constant

\begin{equation}
\alpha=\frac{8 \pi^5k^4}{15h^3c^3} \,.
\label{n}
\end{equation}
If we say the contribution of radiation to the total pressure is a fraction of $(1-\beta)$, we can write

\begin{equation}
p=\frac{1}{1-\beta}\alpha T^4=\frac{1}{\beta}\frac{k}{\mu \textbf{H}}\rho T \,.
\label{o}
\end{equation}
By eliminating $T$ from two equations above, we reach

\begin{equation}
T=\left[\frac{k}{\mu \textbf{H}}\frac{3}{\alpha}\frac{1-\beta}{\beta}\right]^{1/3}\rho^{1/3} \,.
\label{p}
\end{equation}
Substituting $T$ in total pressure, we obtain

\begin{equation}
p=\left[\left(\frac{k}{\mu \textbf{H}}\right)^4\frac{3}{\alpha}\frac{1-\beta}{\beta^4}\right]^{1/3}\rho^{4/3}=K\rho^{4/3} \,,
\label{r}
\end{equation}
where we obtain the polytropic index of $\gamma=4/3$ in \eqref{k} which will be used throughout the whole paper for our fluid constituent of the universe in a polytropic adiabatic process.

\end{document}